\documentclass[jkps,twoside,twocolumn,fleqn,showpacs,showkeys,showdoi]{revtex4}
\usepackage[pdftex]{graphicx}
\usepackage{amssymb}
\usepackage{amsmath}
\usepackage{bm,multirow}
\usepackage{eurosym}
\usepackage{enumerate}
\usepackage{slashbox}
\usepackage{subeqnarray}
\usepackage{undertilde}
\newcommand{\hs}{\hspace{-0.4cm}}
\newcommand{\nn}{\nonumber}

\def\be{\begin{equation}}
\def\ee{\end{equation}}
\def\ba{\begin{eqnarray}}
\def\ea{\end{eqnarray}}
\volumenumber{68} \issuenumber{6} \volumeyear{March 2016}
\startpage{0}
\endpage{0}
\begin{document}
\setcounter{page}{1}

\title[Black Hole Entropy Predictions Without the Immirzi Parameter ]
{Black Hole Entropy Predictions Without the Immirzi Parameter and Hawking Radiation of a Single-partition Black Hole}

 \author{Brian \surname{Kong}}
    \email{kwt0506@gmail.com}
    \affiliation{400 Memorial Dr, Cambridge, MA 02139, USA}

 \author{Youngsub \surname{Yoon}}
    \email{youngsuby@snu.ac.kr}
    \affiliation{Department of Physics and Astronomy, Seoul National University, Seoul
        08826, Korea}

\date[]{Received 4 February 2016, in final form 19 February 2016}

\begin{abstract}
By pointing out an error in the previous derivation of the area
spectrum based on Ashtekar's variables, we suggest a new area
spectrum; instead of the norm of Ashtekar's gravitational electric
field, we show that the norm of our ``new'' gravitational electric
field based on our ``newer'' variables, which we construct in this
paper for this purpose, gives the correct area spectrum. In
particular, our ``newer'' variables are mathematically consistent;
the constraint algebra is closed. Moreover, by using our new area
spectrum, we ``almost correctly'' predict the Bekenstein-Hawking
entropy without having to adjust the Immirzi parameter; we show that
a numerical formula actually yielded $0.997\cdots$, which is very
close to 1, the expected value with the black hole entropy given as
$A/4$. We conjecture that the difference, 0.003, is due to the extra
dimensions that may modify the area spectrum. Then, we derive a
formula for the degeneracy for a single-partition black hole,
\textit{i.e.}, a black hole made of a single unit area, and
explicitly show that our area spectrum correctly reproduces the
degeneracy. Furthermore, by using two totally different methods, we
obtain the proportionality constant ``$C$'' related to the
degeneracy. The first method based on fitting yields 172 $\sim$ 173
while the second method yields 172.87$\cdots$, which strongly
suggest that our area spectrum is on the right track. We also show
that the area spectra based on Ashtekar variables neither reproduce
the degeneracy of single-partition black hole nor yield agreement
for $C$ obtained by using the two methods.
\end{abstract}



\maketitle

\section {Introduction}
According to loop quantum gravity, the eigenvalues of the area
operator admit only discrete values, which means that the area
spectrum is quantized. Ashtekar variables\cite{Ashtekar} are known
to be useful when calculating the area
spectrum\cite{Quantum,area,Area-length}. However, in this paper, we
argue that the area spectrum so obtained is wrong and suggest a new
area spectrum. The problem is not that the Asthekar formulation is
wrong, but the paper\cite{Area-length}, which claimed that the norm
of the gravitational electric field in Ashtekar formulation gives
the area spectrum, suffers from an error; in this paper, we suggest
that the norm of our ``newer'' gravitational electric field gives
the correct area spectrum. To this end, we will propose ``newer''
variables, which look similar to Ashtekar variables, but nonetheless
as different. This similarity enables us to make the constraints in
our canonical quantization look similar to those in the Ashtekar
formulation, making the constraints algebra closed as in the
Ashtekar formulation, while the difference gives the new area
spectrum, as the commutator of the connection and the gravitational
electric field is replaced by something similar, but different.

After obtaining the new area spectrum, we insert the result into a
variation of the formula which can be found in Ref. 5. We obtain
$0.997\cdots$ for the numerical calculation that should, when
naively considered, come out to be 1 if the famous
Bekenstein-Hawking entropy formula ($S=A/4$\cite{Bekenstein_entropy,
Hawking}) is satisfied. The fact that this numerical value we have
obtained is so close to 1 strongly suggests that the new area
spectrum that we will present in this paper is ``almost'' correct.
We conjecture that this difference of 0.003 is due to the extra
dimensions, which may modify the area spectrum. This concludes the
first main argument which is presented from Section II to section
VII.

In the second phase of our paper, we present another convincing
proof that this area spectrum obtained in Section VI is correct. To
this end, we consider the Hawking radiation spectrum. Black holes
are well known to have to radiate photons, the spectrum of which is
given by Planck's law\cite{Hawking}. In 1995, Bekenstein and
Mukhanov tried to derive this spectrum on the assumptions that only
one unit area existed and that all the spectra of the area were
integer multiples of this unit area\cite{Bekenstein}.

In Section VIII and sections following, we will generalize
Bekenstein and Mukhanov's argument in order to apply it to the case
in which there exist not a single unit area, but a multiple number
of unit areas, as we know now. In Section VIII, we will present the
reasoning for why the number of microstates (\textit{i.e.},
degeneracy) for a black hole with area $A$ should be proportional to
$\sqrt{A}(e^{A/4}-1)$ rather than $e^{A/4}$ when $A$ is so small
that the black hole is made out of single partition. This will be
further justified in Section XI.

In Section IX, we will use the area spectrum obtained in Section VI
to numerically prove that the number of microstates corresponding to
the single-partition black hole (\textit{i.e.}, a black hole made of
a single unit area) is, indeed, given by $\sqrt{A}(e^{A/4}-1) dA$
divided by a certain constant that we call $C$, which we find to be
172 $\sim$ 173. In Sections X-XII, by closely following Bekenstein
and Mukhanov's paper\cite{Bekenstein}, we will calculate the
constant $C$, which we defined and obtained in Section VIII by using
a totally different method and show that it gives a value of
172.87$\cdots$ in complete agreement with 172 $\sim$ 173.

In section XIII, we argue that the black hole's degeneracy
(\textit{i.e.}, the exponential of the black hole's entropy) is
given by the naive black-hole degeneracy, recently proposed and
named by one of us, if the area spectrum on the black hole's horizon
is given by the general area spectrum (either in its newer variable
version or in the original Ashtekar variable version) rather than an
isolated horizon framework. In section XIV, we conclude our paper
and suggest some directions for future research. In appendix A, we
will show that the statistical fitting that we performed in Section
IX does not work at all for any of the three area spectra based on
Ashtekar variables. In appendix B, we review the Kaluza-Klein
theory, which we guess to be essential for validating our conjecture
about the effect of an extra-dimension, and the works of Einstein
\textit{et al.} on the Kaluza-Klein theory.

\section {Area is the length of our ``new'' gravitational electric field}

In Ref. 4, Rovelli asserted that area is given by the length of $E$
\textit{i.e.}, the gravitational electric field, where the
definition is given by
\begin{equation}
g g^{ab}=\sum_i E^{ia} E^{ib}.\label{ggab}
\end{equation} (Our convention in this section is as follows: we use $i$, $j$, $k$ for Lorentz indices and $a$, $b$, $c$, $d$, $e$, $f$ for space indices. We denote the Levi-Civita symbol by using a tilde as in $\tilde{\epsilon}_{123}\equiv 1$ while the Levi-Civita tensor is written without tilde as in $\epsilon_{123}\equiv \sqrt g$.) Rovelli obtained this result by considering the area element as follows:
\begin{equation}
E^i(x)=E^{ia}(x)\tilde{\epsilon}_{abc}dx^b \wedge dx^c,\label{Eia}
\end{equation}
where $\tilde{\epsilon}_{123}\equiv 1$ and is called the Levi-Civita
symbol. However, this relation is not correct; a careful
combinatorics factor analysis shows that an extra $\frac 12$ factor
is missing. In other words, we should have
\begin{equation}
E^i(x)=\frac {1}{2!} E^{ia}(x)\tilde{\epsilon}_{abc}dx^b \wedge
dx^c.\label{stillwrong}
\end{equation}
However, Eq. (\ref{stillwrong}) is still wrong, because he
incorrectly used the Levi-Civita symbol (\textit{i.e.},
$\tilde{\epsilon}_{123}\equiv 1$) where he should have used the
Levi-Civita tensor (\textit{i.e.}, $\epsilon_{123}=\sqrt g$), as
$a$, $b$, $c$ are not Lorentz indices, but spacetime indices. If
$a$, $b$, $c$ were Lorentz indices, we could have used
$\tilde{\epsilon}_{123}\equiv 1$, but that was not the case.
Therefore, $E^{ia}(x)$ in Eq. (\ref{stillwrong}) is not appropriate.
For example, in string theory, the action of a string in the
presence of background fields is given as
\begin{eqnarray}
 S &=&  \frac{1}{4\pi\alpha'}\int d^2\sigma \sqrt{g}
[(g^{ab}G_{\mu\nu}(X)+i\epsilon^{ab} B_{\mu\nu}(X)) \nn \\ &&
\qquad\qquad\qquad ~~~ \times \partial_b X^\nu
\partial_a X^\mu +\alpha' R \Phi(X)], ~
\end{eqnarray}
where $\epsilon^{ab}$ is not a Levi-Civita symbol, but a Levi-Civita
tensor.

Given this, contrary to what Rovelli argued, in this section, we
will assert that the area is given by the length of $D$, our ``new''
gravitational electric field, which is defined as follows:
\begin{equation}
g^{ab}=\sum_i D^{ia} D^{ib},\label{before}
\end{equation}
\textit{i.e.}, $D^{ia}=e^{ia}$, the dreibein. To this end, let us
write the area element as follows:
\begin{equation}
E^i(x)=\frac{1}{2!}D^{ia}(x)\epsilon_{abc} dx^b \wedge
dx^c,\label{Dia}
\end{equation}
where $\epsilon_{123}\equiv \sqrt{g}$ and is called the Levi-Civita
tensor, as $a$, $b$, $c$ are spacetime indices. Another way of
seeing that this is reasonable is that the formula for the hodge
dual always includes $\sqrt{g}$. In other words,
\begin{equation}
(dx_a)^*=\frac{1}{2!}\epsilon_{abc} dx^b \wedge dx^c.
\end{equation}

Now, comparing Eqs.~(\ref{stillwrong}) and (\ref{Dia}), we have
\begin{equation}
D^{ia}\sqrt{g}=E^{ia}
\end{equation}

Therefore, from Eq.~(\ref{ggab}), we conclude that
Eq.~(\ref{before}) holds. In conclusion, we have
\begin{equation}
E^i(x)=\frac{1}{2!}e^{ia}(x)\epsilon_{abc} dx^b \wedge dx^c.
\end{equation}
Notice that each term, \textit{i.e.}, $e^{Ia}$, $\epsilon_{abc}$ in
the multiplication is a fully covariant tensor whereas such a
statement cannot hold true for Eq. (\ref{Eia}).

\section{Review of Ashtekar variables}
In this section, we will review the Ashtekar variables closely
following Ref.~2. This step is important because our construction of
our ``newer'' variables will closely follow this construction. The
convention in this section is as follows. The Lorentz indices $i$,
$j$, $k$ take 1, 2, and 3 for their values, and the Lorentz indices
$I$, $J$, $K$ take 0, 1, 2, and 3 as their values. Also, we use
Greek letters for spacetime indices and $a$, $b$, $c$ for space
indices.

\subsection{Vierbein Formalism and Palatini Action}
Let us consider the vierbein formalism of general relativity as
follows:
\begin{eqnarray}
&&g_{\mu\nu}=e^I_{\mu} e^J_{\nu} \eta_{IJ},\\
&&e^I(x)=e^I_{\mu}(x) dx^{\mu},\\
&&\omega^I {}_{J} (x)=\omega^I_{\mu J}(x) dx^{\mu},\\
&&\omega^{IJ}=-\omega^{JI},\\
&&Du^I=du^I+\omega^I {}_{J} \wedge u^J,\\
&&R^I {}_{J}=d\omega^I {}_{J}+\omega^{I} {}_{K} \wedge \omega^K
{}_{J}.
\end{eqnarray}
Given this, the Palatini action is given by
\begin{eqnarray}
S&=&\frac{1}{16\pi G}\int d^4x \sqrt{-g} R\nn\\&=&\frac{1}{16\pi
G}\int \epsilon_{IJKL}\frac{1}{2} e^I\wedge e^J \wedge
R^{KL}\label{Palatiniaction}
\end{eqnarray}

\subsection{Self-dual Formalism and Plebanski Formalism}
Given Eq.~(\ref{Palatiniaction}), let us introduce the self-dual
formalism. To do so, we introduce the self-dual projector $P^i_{JK}$
as follows:
\begin{eqnarray}
&& P^i_{jk}=\frac{1}{2} \tilde \epsilon^i {}_{jk} ,\nonumber\\
&& P^i_{0j}=-P^i_{j0}=\frac{i}{2}\delta^i_j. \label{P}
\end{eqnarray}
Then the complex SO(3) connection is given as
\begin{equation}
A^{k}_{\mu}dx^{\mu}=P^k_{I J} \omega^{I J}_{\mu}
dx^{\mu},\label{complex}
\end{equation}
which implies,
\begin{eqnarray}
&& A^1=\omega^{23}+i\omega^{01},\\
&& A^2=\omega^{31}+i\omega^{02},\\
&& A^3=\omega^{12}+i\omega^{03}.
\end{eqnarray}
As an aside, in the presence of the Immirzi parameter, we have
\begin{equation}
A^1=\omega^{23}+\gamma \omega^{01}
\end{equation}
and similarly for the other components, where $\gamma$ is the
Immirzi parameter, which is real.

Using this formalism, one can write the Einstein-Hilbert action as
\begin{equation}
S=\frac{2}{16 \pi G}\int-i P_{kIJ} e^I \wedge e^J \wedge
F^k,\label{PeeF}
\end{equation}
where $F^k$ is given by
\begin{equation}
F^k=dA^k + \tilde \epsilon^k{}_{lm} A^l A^m.
\end{equation}
Let's prove this:
\begin{widetext}
\begin{eqnarray}
F^1&=&dA^1+A^2\wedge A^3\\
&=&d\omega^{23}+\omega^{31}\wedge\omega^{12}-\omega^{02}\wedge\omega^{03}
+i(d\omega^{01}+\omega^{02}\wedge\omega^{12}+\omega^{31}\wedge\omega^{03})\\
&=&d\omega^{23}-\omega^{21}\wedge\omega^{13}+\omega^{20}\wedge\omega^{03}+i(d\omega^{01}-\omega^{02}\wedge\omega^{21}-\omega^{03}\wedge\omega^{31})\\
&=&d\omega^{23}+\omega^{2}{}_{1}\wedge\omega^{13}+\omega^{2}{}_{0}\wedge\omega^{03}+i(d\omega^{01}+\omega^{0}{}_{2}\wedge\omega^{21}+\omega^{0}{}_{3}\wedge\omega^{31})\\
&=&R^{23}+iR^{01}.
\end{eqnarray}
\end{widetext}
Similarly, we have:
\begin{equation}
F^i=\frac 12 \tilde \epsilon^i{}_{jk}R^{jk}+iR^{0i}.
\end{equation}
Now the Eq.~(\ref{PeeF}) becomes

\begin{eqnarray}
&&\hs\hspace{-0.5cm} S=\frac{2}{16\pi G}\int (e^0\wedge e_i\wedge F^i-i\frac 12\tilde \epsilon_{ijk} e^j\wedge e^k \wedge F^i)\label{Plebanskifirst}\\
&&\hs\hspace{-0.5cm} =\frac{2}{16\pi G}\int(\frac 12 \tilde \epsilon_{ijk}e^0\wedge e^i\wedge R^{jk}+\frac 12 \tilde \epsilon_{ijk}e^j\wedge e^k\wedge R^{0i})\label{Plebanskisecond}\\
&&\hs  +\frac{2i}{16\pi G}\int (e^0\wedge e^i \wedge R^0{}_{i}-\frac
12 e_l\wedge e^m\wedge R^l{}_{m})\label{Plebanskithird}\\
&&\nn
\end{eqnarray}
where from Eqs. (\ref{Plebanskifirst}) to (\ref{Plebanskithird}),
$\tilde\epsilon_{ijk}\tilde
\epsilon^{ilm}=\delta^l_j\delta^m_k-\delta^l_k\delta^m_j$ was used.
Notice also that the third line vanishes from the Bianchi identity
$R^i{}_{j}\wedge e^j=0$. Furthermore, one can easily see that the
second line is equal to the Palatini action in Eq.
(\ref{Palatiniaction}). Thus, we, indeed, see that Eq. (\ref{PeeF})
is equal to the Palatini action. Given this, let us rewrite
(\ref{PeeF}) in so-called Plebanski formalism, by defining the
Plebanski self-dual two-form as follows:
\begin{equation}
\Sigma^k=P^k_{IJ} e^I \wedge e^J;
\end{equation}
that is,
\begin{eqnarray}
&& \Sigma^1=e^2 \wedge e^3 + i e^0 \wedge e^1\label{Sigma1e0}\\
&& \Sigma^2=e^3 \wedge e^1 + i e^0 \wedge e^2\label{Sigma2e0}\\
&& \Sigma^3=e^1 \wedge e^2 + i e^0 \wedge e^3\label{Sigma3e0}
\end{eqnarray}
Therefore, we can write the Einstein-Hilbert action in the Plebanski
formalism as follows:
\begin{equation}
S=\frac{-2i}{16\pi G}\int \Sigma_k \wedge F^k
\end{equation}
Given this, we now turn our attention to the Ashtekar formalism.

\subsection{Ashtekar Formalism}
Consider a solution $(e^I_\mu(x), A^I_\mu(x))$ of the Einstein
equations. Choose a 3d surface
$\sigma:\overrightarrow{\tau}=(\tau^a)\rightarrow
x^\mu(\overrightarrow{\tau})$ without boundaries in coordinate
space. The four-dimensional forms $A^I$, $\Sigma^I$ and $e^I$ induce
the following three-dimensional forms:
\begin{eqnarray}
&& A^I(\overrightarrow{\tau})=A^I_a(\overrightarrow{\tau}) d\tau^a, \\
&& \Sigma_I(\overrightarrow{\tau})= \frac 12 \Sigma_{Iab}(\overrightarrow{\tau}) d\tau^a \wedge d\tau^b, \\
&& e^I(\overrightarrow{\tau})=e^I_a(\overrightarrow{\tau}) d\tau^a,
\end{eqnarray}

Let us write $e^I=(e^0, e^i)$, and choose a gauge in which
\begin{equation}
e^0=0.
\end{equation}
In this gauge, from Eqs.~(\ref{Sigma1e0}), (\ref{Sigma2e0}), and
(\ref{Sigma3e0}), we have
\begin{eqnarray}
&& \Sigma^1=e^2 \wedge e^3, \nonumber\\
&& \Sigma^2=e^3 \wedge e^1, \nonumber\\
&& \Sigma^3=e^1 \wedge e^2,
\end{eqnarray}
which is, indeed, the area two-form. Now, the area operator can be
re-expressed as
\begin{eqnarray}
&& \Sigma^i=E^i,\\
&& \frac 12 \Sigma^i_{bc} dx^b\wedge dx^c=\frac 12 E^{id}\tilde{\epsilon}_{dbc}dx^b \wedge dx^c,\\
&& \Sigma^i_{bc}= E^{id}\tilde{\epsilon}_{dbc}.\\
&&\nn
\end{eqnarray}
Multiplying both sides by $\tilde{\epsilon}^{abc}$, we obtain
\begin{equation}
E^{ia}=\frac 12 \tilde{\epsilon}^{abc}\Sigma^i_{bc}.\label{EIa}
\end{equation}
Furthermore, from
\begin{equation}
\Sigma_i=E_i=\frac 12 \tilde {\epsilon}_{ijk}e^j\wedge e^k=\frac 12
\Sigma^i_{bc} dx^b\wedge dx^c,
\end{equation}
we have
\begin{equation}
\Sigma_{ibc}=\tilde{\epsilon}_{ijk}e^j_b e^k_c,
\end{equation}
which, from Eq. (\ref{EIa}), implies
\begin{equation}
E^a_i=\frac 12 \tilde{\epsilon}^{abc}\tilde{\epsilon}_{ijk}e^j_b
e^k_c.
\end{equation}
Given this, we can now obtain the action in terms of the Ashtekar
variables:
\begin{widetext}
\begin{eqnarray}
S&=&\frac{-2i}{16\pi G}\int \Sigma_i \wedge F^i=\frac{-2i}{16\pi G}\int \frac 14 \Sigma_{i\mu\nu}F^i_{\rho\sigma}\tilde{\epsilon}^{\mu\nu\rho\sigma}d^4 x\nonumber\\
&=&\frac{-2i}{16\pi G}\int \frac 12(\Sigma_{iab} F^i_{c0}+\Sigma_{i0a}F^i_{bc})\tilde{\epsilon}^{abc}d^4x\nonumber\\
&=&\frac{-i}{8\pi G}\int (E^c_i(\partial_0 A^i_c-\partial_c A^i_0 + \tilde \epsilon^i{}_{jk}A^j_0A^k_c)+P_{iJK} e^J_a e^K_0 F^i_{bc} \tilde{\epsilon}^{abc})d^4x\label{Ashtekarthird}\\
&=&\frac{-i}{8\pi G}\int(E^c_i\dot{A}^i_c +A^i_0 D_c E^c_i +\frac{1}{2} (\tilde \epsilon^i{}_{jk} e^j_a e^k_0+ i e^0_0 e^i_a)F_{ibc}\tilde{\epsilon}^{abc})d^4x\label{Ashtekarfourth}\\
&=&\frac{-i}{8\pi G}\int (E^c_i\dot{A}^i_c +\lambda^i_0 (D_c E^c_i)
+ \lambda^b (E^a_i F^i_{ab}) +\lambda (\tilde
\epsilon^{jk}{}_{i}E^a_j E^b_k F^{i}_{ab}))d^4x,\label{Ashtekar}
\end{eqnarray}
\end{widetext}
where from Eq.~(\ref{Ashtekarthird}) to Eq.~(\ref{Ashtekarfourth})
we used integration by parts, \textit{i.e.}, the total derivatives
do not contribute to the integration. From the above action, the
following Poisson bracket can be easily read:
\begin{equation}
\{A^i_a(\overrightarrow{\tau}), E^b_j(\overrightarrow{\tau}')\}=(i)
\delta^i_j \delta^b_a
\delta^3(\overrightarrow{\tau},\overrightarrow{\tau}')\label{Poisson}
\end{equation}
where we have set $8\pi G=1$. Also from the action, the following
constraints can be easily read:
\begin{eqnarray}
&& D_c E^c_i=0, \nonumber\\
&& E_i^a F^i_{ab}=0, \nonumber\\
&& \tilde \epsilon^{jk}{}_i E^a_j E^b_k
F^{i}_{ab}=0,\label{Constraints}
\end{eqnarray}

Given the above, let us use the following notation:
\begin{eqnarray}
&& \mathcal G(N_i)=\int d^3x N_i D_c E^c_i,\nonumber\\
&& C(\vec N)=\int d^3 x N^bE_i^a F^i_{ab}-\mathcal G(N^aA_a^i),\nonumber\\
&& \mathcal H(\utilde N)=\int d^3x \utilde N
\tilde\epsilon^{ij}{}_{k}E^a_i E^b_j F^k_{ab}.
\end{eqnarray}
Then, using Eq.~(\ref{Poisson}), one can show that the constraint
algebra is closed as follows:
\begin{eqnarray}
\{\mathcal G(N_i), \mathcal G(N_j)\}&=&\mathcal G([N_i,N_j]),\nonumber\\
\{C(\vec N), C(\vec M)\}&=&C(\mathcal L_{\vec M}\vec N),\nonumber\\
\{C(\vec N),\mathcal G(N_i)\}&=&\mathcal G(\mathcal L_{\vec N}N_i),\nonumber\\
\{C(\vec N), \mathcal H(\utilde M)\}&=&\mathcal H(\mathcal L_{\vec N}\utilde M),\nonumber\\
\{\mathcal G(N_i),\mathcal H(\utilde N)\}&=&0,\nonumber\\
\{\mathcal H(\utilde N),\mathcal H(\utilde M)\}&=&C(\vec K)-\mathcal
G(A^i_a K^a),\label{constraintAshtekar}
\end{eqnarray}
where
\begin{equation}
K^a=2E^a_iE^b_i(\utilde N\partial_a \utilde M-\utilde
M\partial_a\utilde N).
\end{equation}

\section{Spin connection in pure spacetime indices}
In this section, we will introduce the spin connection in the
spacetime variables, which is slightly different from the generic
spin connection. A generic spin connection is given as follows:
\begin{equation}
\omega^{I} {}_{J}=\omega^I_{\mu J} dx^{\mu},
\end{equation}
where $I$ and $J$ are Lorentz indices, and $\mu$ is the spacetime
index. Given this, let us construct the spin connection in pure
spacetime indices.

In the differential form formalism of general relativity, the
covariant derivative is defined as\cite{Ryder}.
\begin{equation}
\nabla V=(dV^{\alpha}+\omega^{\alpha} {}_{\beta}
V^{\beta})e_{\alpha}, \label{differentialformcovariant}
\end{equation}
while in the metric formalism, the covariant derivative is defined
as
\begin{equation}
\nabla_{\mu}V^{\alpha}=\partial_{\mu}V^{\alpha}+\Gamma^{\alpha}_{\mu
\beta} V^{\beta}. \label{metriccovariant}
\end{equation}
Again, in the case of the differential form formalism, we have the
following for the Riemann tensor:
\begin{equation}
R^{\alpha} {}_{\beta}=d\omega^{\alpha} {}_{\beta}+\omega^{\alpha}
{}_{\kappa}\wedge\omega^{\kappa} {}_{\beta}=R^{\alpha}_{\beta \mu
\nu} dx^{\mu}\wedge dx^{\nu}.
\end{equation}
By an explicit calculation, we obtain
\begin{equation}
R^{\alpha}_{\beta\mu\nu}=\omega^{\alpha}_{\beta\nu,
\mu}-\omega^{\alpha}_{\beta\mu, \nu}+\omega^{\alpha}_{\kappa
\mu}\omega^{\kappa}_{\beta \nu}-\omega^{\alpha}_{\kappa
\nu}\omega^{\kappa}_{\beta\mu}\label{differentialformRiemann}.
\end{equation}
In the case of the metric formalism, we have the following for the
Riemann tensor:
\begin{equation}
R^{\alpha}_{\beta\mu\nu}=\Gamma^{\alpha}_{\beta\nu,
\mu}-\Gamma^{\alpha}_{\beta\mu, \nu}+\Gamma^{\alpha}_{\kappa
\mu}\Gamma^\kappa_{\beta \nu}-\Gamma^{\alpha}_{\kappa
\nu}\Gamma^{\kappa}_{\beta\mu} \label{metricRiemann}.
\end{equation}

Therefore, we easily see that we get Eqs.~(\ref{metriccovariant})
and (\ref{metricRiemann}) if we replace $\omega$ by $\Gamma$ in
eqs.~(\ref{differentialformcovariant}) and
(\ref{differentialformRiemann}). Indeed, in Ref.~9, the connection
one-form is defined as
\begin{equation}
\omega^\kappa {}_\lambda=\Gamma^\kappa_{\lambda\mu} dx^{\mu}.
\end{equation}
This is the spin connection in pure spacetime indices as advertised.
Notice also the following, which can be derived from the fact that
the covariant derivative of metric vanishes:
\begin{equation}
dg^{\mu\nu}=\omega^{\mu\nu}+\omega^{\nu\mu}.\label{dgmunu}
\end{equation}
We will use this spin connection in the next section.

\section{Our ``newer'' variables}
In this section, we will rewrite the Einstein-Hilbert action in
terms of our ``newer'' variables instead of Ashtekar variables,
which he called ``new variables''\cite{Ashtekar}. To this end, we
will define a ``tautological'' vierbein which has spacetime indices
for both upper and lower indices as follows:
\begin{eqnarray}
&& g_{\alpha \beta}=g_{\mu \nu} \tilde e^\mu_\alpha \tilde e^\nu_\beta,\\
&& \tilde e^\mu=\tilde e^\mu_\alpha dx^\alpha,
\end{eqnarray}
Now, analogous to Eq.~(\ref{complex}), we will define our newer
complex $SO(3)$ connection $\tilde{A}$ as follows:
\begin{equation}
\tilde{A}^{i}_{\mu}dx^{\mu}=\tilde P^i_{\alpha\beta}
\omega^{\alpha\beta}_{\mu} dx^{\mu}, \label{Aia}
\end{equation}
where unlike in Section III, $i$, $j$ and $k$ are now space indices.
Here, the self-dual projector is given by
\begin{equation}
\tilde P^i_{jk}=\frac 12 \epsilon^i{}_{jk},\qquad \tilde
P^i_{oj}=-\tilde P^i_{j0}=\frac i2 \delta^i_j.
\end{equation}
(Notice that $\epsilon$ here is a Levi-Civita tensor, not a
Levi-Civita symbol.) From this, we can also construct the curvature
two-form $\tilde{F}^i$ as follows:
\begin{equation}
\tilde{F}^i=d\tilde{A}^i + \epsilon^i{}_{jk} \tilde{A}^j
\tilde{A}^k,
\end{equation}
where $\epsilon$ here again is a Levi-Civita tensor. Now, let us
impose a gauge condition that the metric be diagonal. Then,
Eq.~(\ref{dgmunu}) suggests that $\omega^{\mu\nu}$ is antisymmetric
if $\mu\neq\nu$. With this condition, we can write the
Einstein-Hilbert action as follows:
\begin{eqnarray}
&&S=\frac{2}{16 \pi G}\int-i \tilde P_{i\mu\nu} \tilde e^\mu \wedge
\tilde e^\nu \wedge \tilde{F}^i.\label{newS}\\
&&\nn\\
&&\nn
\end{eqnarray}
Let us show this. We have
\begin{eqnarray}
A^1&=&g^{11}\sqrt{g}\left(\frac 12 (\omega^{23}-\omega^{32})\right)+\frac 12 i (\omega^{01}-\omega^{10})\nonumber\\
&=&g^{11}\sqrt{g}\omega^{23}+i\omega^{01},
\end{eqnarray}
and similarly for $A^2$ and $A^3$. Now, notice the following:
\begin{widetext}
\begin{eqnarray}
R^{23}&=&\frac 12(R^{23}-R^{32})\\
&=&\frac 12(d\omega^{23}+\omega^2{}_1\wedge\omega^{13}+\omega^2{}_0\wedge\omega^{03}-d\omega^{32}-\omega^3{}_1\wedge\omega^{12}-\omega^3{}_0\wedge\omega^{02})\nonumber\\
&&+\frac 12 (\omega^2{}_2\wedge \omega^{23}-\omega^3{}_2\wedge\omega^{22}+\omega^2{}_3\wedge \omega^{33}-\omega^3{}_3\wedge \omega^{32})\label{vanishing}\\
&=&d\omega^{23}+\omega^2{}_1\wedge\omega^{13}+\omega^2{}_0\wedge\omega^{03},
\end{eqnarray}\end{widetext}
where Eq.~(\ref{vanishing}) vanishes due to Eq.~(\ref{dgmunu}).
Then, we have
\begin{eqnarray}
F^1&=&dA^1-g^{11}\sqrt g A^2\wedge A^3\\
&=&g^{11}\sqrt g R^{23}+i R^{01},
\end{eqnarray}
where $g_{00}=g^{00}=-1$ is used.

Similarly, we have
\begin{equation}
F^i=\frac 12 \epsilon^i_{jk}R^{jk}+i R^{0i}.
\end{equation}
Taking the same step as Eqs.~(\ref{Plebanskifirst}),
(\ref{Plebanskisecond}) and (\ref{Plebanskithird}), we obtain
\begin{equation}
S=\frac{1}{16\pi G}\int \epsilon_{\mu\nu\rho\sigma}\frac 12 \tilde
e^\mu \wedge \tilde e^\nu \wedge R^{\rho\sigma},
\end{equation}
where $\epsilon$ here is a Levi-Civita tensor and we can use $\tilde
e^\mu_\alpha=\delta^\mu_\alpha$ to prove that this is equal to the
Einstein-Hilbert action.

Given this and using the definition
\begin{equation}
\tilde{\Sigma}^i=\tilde P^i_{\mu\nu} \tilde e^\mu \wedge \tilde
e^\nu,
\end{equation}
from Eq.~(\ref{newS}), we can write the ``newer'' Plebanski action
as follows:

\begin{equation}
S=\frac{-2i}{16\pi G}\int \tilde{\Sigma}_i \wedge \tilde{F}^i.
\end{equation}
Now, everything in the Section III follows naturally. We have
\begin{eqnarray}
&& \tilde{A}^i(\overrightarrow{\tau})=\tilde{A}^i_a(\overrightarrow{\tau}) d\tau^a, \\
&& \tilde{\Sigma}^i(\overrightarrow{\tau})=\frac 12 \tilde{\Sigma}^i_{ab}(\overrightarrow{\tau}) d\tau^a \wedge d\tau^b, \\
&& \tilde e^i(\overrightarrow{\tau})=\tilde
e^i_a(\overrightarrow{\tau}) d\tau^a,\\
&&\nn
\end{eqnarray}
and the following definition for $\tilde{E}^{ia}$:
\begin{equation}
\tilde{E}^{ia}=\frac 12
\tilde{\epsilon}^{abc}\tilde{\Sigma}^i_{bc}\label{Eai}
\end{equation}

Given this, let us rewrite the action. Equation (\ref{Ashtekar})
becomes
\begin{widetext}
\begin{eqnarray}
S&=&\frac{-2i}{16\pi G}\int \tilde{\Sigma}^i \wedge \tilde{F}_i=\frac{-2i}{16\pi G}\int \frac 14\tilde {\Sigma}^i_{\mu\nu}\tilde{F}_{i\rho\sigma}\tilde{\epsilon}^{\mu\nu\rho\sigma}d^4 x\nonumber\\
&=&\frac{-i}{8\pi G}\int (\tilde{E}^{ic}\dot{\tilde{A}}_{ic}
+\lambda^i_0 (D_c \tilde{E}^c_i) + \lambda^b (\tilde{E}^a_i
\tilde{F}^i_{ab}) +\lambda (\epsilon^{jk}{}_i\tilde{E}^a_j
\tilde{E}^b_k \tilde{F}^{i}_{ab}))d^4x.\label{newervariables}
\end{eqnarray}\end{widetext}
Now, notice the following, which is satisfied when the metric is
diagonal:
\begin{equation}
\tilde{A}_i=\tilde{A}_{i\mu} dx^\mu=(\frac{1}{2} \epsilon^k_{ij}
\Gamma^j_{\mu k}+i \Gamma^0_{\mu i})dx^\mu,
\end{equation}
which implies
\begin{equation}
\dot{\tilde{A}}_{ic}=i(\partial_0 \Gamma^o_{ci})+\cdots
\end{equation}
Given this, notice also the following:
\begin{eqnarray}
S &=& \frac{1}{16\pi G}\int \sqrt{g} R d^4x\nn \\
&=& \frac{1}{16\pi G}\int \sqrt{g} (g^{ci} \partial_0
\Gamma^0_{ci}+\cdots) d^4x.
\end{eqnarray}
One can check that
\begin{equation}
\frac{\delta (\sqrt g R)}{\delta(\partial_0 \Gamma^0_{ci}=\partial_0
\Gamma^0_{ic})}=\sqrt{g} (g^{ic}+g^{ci})=2\sqrt g g^{ic}.
\end{equation}
This relation holds in the case $c=i$, for example,
\begin{equation}
\frac{\delta (\sqrt g
(R=R^{01}{}_{01}+R^{10}{}_{10}+\cdots))}{\delta(\partial_0
\Gamma^0_{11})}=2\sqrt{g} g^{11}
\end{equation}
Therefore, we obtain
\begin{equation}
\tilde{E}^{ic}=\sqrt{g} g^{ic}.
\end{equation}

Now, that one can always choose the gauge $g=1$ is well known, as
Einstein noted in his paper on general relativity\cite{Einstein}. In
this gauge, we have
\begin{equation}
\tilde{E}^{ic}= g^{ic}.\label{tildeEai}
\end{equation}

On the other hand, Eq.~(\ref{Poisson}) becomes
\begin{equation}
\{\tilde{A}_{ic}(\overrightarrow{\tau}),
\tilde{E}^{jb}(\overrightarrow{\tau}')\}=(i) \delta^j_i \delta^b_c
\delta^3(\overrightarrow{\tau},\overrightarrow{\tau}'),\label{newPoisson}
\end{equation}
and the constraints in Eq.~(\ref{Constraints}) become
\begin{eqnarray}
&& D_c \tilde{E}^c_i=0, \nonumber\\
&& \tilde{E}_i^a \tilde{F}^i_{ab}=0, \nonumber\\
&& \epsilon^{jk}{}_i\tilde{E}^a_j \tilde{E}^b_k
\tilde{F}^{i}_{ab}=0,
\end{eqnarray}

Given this, let us use the following notation:
\begin{eqnarray}
&&\tilde{\mathcal G}(\tilde N_i)=\int d^3x \tilde N_i D_c \tilde E^c_i\nonumber\\
&&\tilde{C}(\vec{\tilde N})=\int d^3 x \tilde N^b\tilde E_i^a \tilde F^i_{ab}-\tilde{\mathcal G}(\tilde N^a\tilde A_a^i)\nonumber\\
&&\tilde{\mathcal H}(\tilde{\utilde N})=\int d^3x \tilde{\utilde N}
\epsilon^{ij}_{~~k}\tilde E^a_i \tilde E^b_j \tilde F^k_{ab}
\end{eqnarray}
Then, exactly as in section III, Eq.~(\ref{newPoisson}) implies that
the constraint algebra is closed:
\begin{eqnarray}
\{\tilde{\mathcal G}(\tilde N_i), \tilde{\mathcal G}(\tilde N_j)\}&=&\tilde{\mathcal G}([\tilde N_i,\tilde N_j]),\nonumber\\
\{\tilde C(\vec{\tilde N}), C(\vec{\tilde M})\}&=&\tilde C(\tilde{\mathcal L}_{\vec{\tilde M}}\vec {\tilde N}),\nonumber\\
\{\tilde C(\vec {\tilde N}),\tilde{\mathcal G}(\tilde N_i)\}&=&\tilde{\mathcal G}(\tilde{\mathcal L}_{\vec{\tilde N} }\tilde N_i),\nonumber\\
\{\tilde C(\vec N), \tilde{\mathcal H}(\tilde{\utilde M})\}&=&\tilde{\mathcal H}(\tilde{\mathcal L}_{\vec{\tilde N}}\tilde{\utilde M}),\nonumber\\
\{\tilde{\mathcal G}(\tilde N_I),\tilde{\mathcal H}(\tilde{\utilde N})\}&=&0,\nonumber\\
\{\tilde{\mathcal H}(\tilde{\utilde N}),\tilde{\mathcal
H}(\tilde{\utilde M})\}&=&\tilde C(\vec{\tilde K})-\tilde{\mathcal
G}(\tilde A^i_a \tilde K^a),\\
&&\nn
\end{eqnarray}
where
\begin{equation}
\tilde K^a=2\tilde E^a_i\tilde E^b_i(\tilde{\utilde N}\partial_a
\tilde{\utilde M}-\tilde{\utilde M}\partial_a\tilde{\utilde N}).
\end{equation}

Therefore, the delicate features of the constraints in the Ashtekar
formalism are preserved in our ``newer'' variable formalism, as the
untilded variables are just replaced with the tilded variables.
Plugging Eqs.~(\ref{before}) and (\ref{tildeEai}) into
(\ref{newPoisson}), we get
\begin{equation}
\{\tilde{A}_{ic}(\overrightarrow{\tau}), \sum_M
D^{Mj}(\overrightarrow{\tau}')D^{Mb}(\overrightarrow{\tau}')\}=(i)
\delta^j_i\delta^b_c
\delta^3(\overrightarrow{\tau},\overrightarrow{\tau}').\label{newerPoisson}
\end{equation}

\section{The spectrum of area}

The fact that our formula, Eq.~(\ref{newerPoisson}), contains two
$D$s instead of one $E$ as in Eq.~(\ref{Poisson}) in the commutator
has a far reaching consequence in calculating the spectrum of area.
As explained earlier, according to the loop quantum gravity based on
Ashtekar's variable, the spectrum of area is calculated to be the
eigenvalues of $\sqrt{E^{a}_{i}E^{bi}}$\cite{Quantum, area,
Area-length}. In Ashtekar's and Rovelli's theory, as $E$ is the
conjugate momentum of $A$, to calculate the previous formula, we
have to replace $E$s by derivatives with respect to $A$s, the
connections. Explicitly, this means that\cite{Quantum}:
\begin{equation}
E^{a}_{i}\Psi_{s}(A)=-i \frac{\delta}{\delta
A^{i}_{a}}\Psi_{s}(A).\label{psisA}
\end{equation}
Here $\Psi_{s}(A)$ is the spin network state.

However, considering Eq. (\ref{newerPoisson}), in this case, we
observe that taking the derivative with respect to $A$ brings down
two $D$s instead of one $E$ as in Eq. (\ref{psisA}) in Rovelli's
theory when taking the derivative with respect to $A$. This means
the following:
\begin{equation}
-i \frac{\delta}{\delta \tilde{A}_{ia}}\Psi_{s}(\tilde{A})=D^{i}_{M}
D^{Ma}\Psi_{s}(\tilde{A}),\label{DD}
\end{equation}
where we have ignored the additional $i$ factor on the right-hand
side of Eq.~(\ref{newerPoisson}). This needs justification, but we
failed to find one. Comparing the above formula with
Eq.~(\ref{psisA}), we can understand that the eigenvalues of the
$D$s of loop quantum gravity based on our ``newer'' variables are
the square roots of the eigenvalues of $E$s (gravitational electric
field) in loop quantum gravity based on Ashtekar's variables. Thus,
we conclude that the area spectra obtained by using our theory are
the square roots of those of the Ashtekar's variables. Therefore, to
obtain the area spectrum predicted by our theory, we first need to
look at the general area spectrum predicted by Ashtekar's and
Rovelli's theory because all we need to do is to take its square
root. To this end, instead of writing down the derivation of the
general area spectrum obtained by using Ashtekar's variable theory,
we simply quote the result\cite{Tanaka, complete}:
\begin{equation}
A=4\pi\gamma\sum_{i}
\sqrt{2j^{u}_{i}(j^{u}_{i}+1)+2j^{d}_{i}(j^{d}_{i}+1)-j^{t}_{i}(j^{t}_{i}+1)},~~
\end{equation}
where $\gamma$ is the Immirzi parameter and $j^{u}_{i}$,
$j^{d}_{i}$, $j^{t}_{i}$ are quantum numbers that satisfy the
constraints found in Ref. 5; \textit{i.e.}, $j^{u}_{i}$, $j^{d}_{i}$
are non-negative half-integers, $j^{t}_{i}$ is a non-negative
integer, and the sum of these three numbers should be an integer and
any sum of the two numbers in this set of three numbers is bigger
than, or equal to, the other number. In particular, the authors of
Ref. 5 note that the condition that $j^{t}_{i}$ is an integer is
``motivated by the ABCK framework where the ``classical horizon'' is
described by a U(1) connection.''\cite{Tanaka, ABCK}. This means
that $j^{t}_{i}$ is an integer because we are considering the case
that the area spectrum lies on black hole's horizon. We would like
to note that in generic cases, this restriction vanishes, and
$j^{t}_{i}$ can be a half integer. Now, the area spectrum of our
theory is the following, as it is the square root of Ashtekar's
variable theory:
\begin{equation}
A=8\pi\sum_{i}
\sqrt{\frac{1}{2}\sqrt{2j^{u}_{i}(j^{u}_{i}+1)+2j^{d}_{i}(j^{d}_{i}+1)-j^{t}_{i}(j^{t}_{i}+1)}}.\label{myformula}
\end{equation}

\section{Black hole entropy calculation}

This section closely follows Ref. 5. To understand the formula that
can check whether the black hole's entropy is $A/4$, we reconsider
the ``simplified area spectrum,'' with the following number of
states $N(A)$:
\begin{equation}
\nn
\end{equation}
\begin{eqnarray}
&&\hs\hs N(A):\nn\\
&&\hs\hs =\left\{(j_{1}, \cdots,j_{n})|0\neq j_{i}\in
\frac{\mathbb{N}}{2},
\sum_{i}\sqrt{j_{i}(j_{i}+1)}=\frac{A}{8\pi\gamma}\right\},\nn\\&&
\end{eqnarray}

We derive a recursion relation to obtain the value of $N(A)$. When
we consider $(j_{1}, \cdots,j_{n})\in N(A-a_{1/2})$, we obtain
$(j_{1}, \cdots,j_{n},\frac{1}{2})\in N(A)$, where $a_{1/2}$ is the
minimum area where only one $j=1/2$ edge contributes to the area
eigenvalue; \textit{i.e.},
$a_{1/2}=8\pi\gamma\sqrt{\frac{1}{2}(\frac{1}{2}+1)}=4\pi\gamma\sqrt{3}$.
Likewise, for any eigenvalue $a_{j_{x}}(0< a_{j_{x}}\leq A)$ of the
area operator, we have
\begin{equation}
(j_{1}, \cdots,j_{n})\in N(A-a_{j_{x}})\Longrightarrow (j_{1},
\cdots,j_{n},j_{x})\in N(A).
\end{equation}
Then, an important point is that if we consider all $0<
a_{j_{x}}\leq A$ and $(j_{1}, \cdots,j_{n})\in N(A-a_{j_{x}})$, $
(j_{1}, \cdots,j_{n},j_{x})$ form the entire set $N(A)$. Thus, we
obtain
\begin{equation}
N(A)=\sum_{j} N(A-8\pi\gamma\sqrt{j(j+1)}).
\end{equation}

By using $N(A)=\exp(A/4)$, one can determine whether the above
formula satisfies the Bekenstein-Hawking entropy formula. If the
Bekenstein-Hawking entropy is satisfied, the following should be
satisfied\cite{Meissner}:
\begin{equation}
1=\sum_{j} \exp(-8\pi\gamma\sqrt{j(j+1)}/4).\label{section6ashtekar}
\end{equation}
Now, we can simply generalize the above formula to the case of the
general area spectrum. First of all, for convenience, Eq.
(\ref{myformula}) is written as
\begin{widetext}
\begin{eqnarray}
&&N(A) :=
\left\{(j^u_1,j^d_1,j^t_1,\cdots,j^u_n,j^d_n,j^t_n)|j^u_i, j^d_i
 \in \frac{\mathbb{N}}{2},\ \ j^t_i \in \mathbb{N},
 \ \ j^u_i+j^d_i+j^t_i\in\mathbb{N},\ j_i^1 \leqq j_i^2 + j_i^3
 \right.  \nonumber  \\\quad
&& \qquad\qquad \left.
\sum_i\sqrt{\frac{1}{2}\sqrt{2j_i^u(j_i^u+1)+2j_i^d(j_i^d+1)-j_i^t(j_i^t+1)}}
=
 \frac{A}{8\pi} \right\}\ .
\end{eqnarray}
\end{widetext}
Moreover, we have to consider some subtleties in counting the number
of state. The authors of Ref. 5 consider ``the proposal that we
should count not only $j$ but also $m=-j,-j+1,\cdots,j-1,j$ freedom
based on the ABCK framework\cite{ABCK}.'' Then, they go on to claim
that counting only $m$ related to $j^{u}$ and $j^{d}$ ``is
reasonable from the point of view of the entanglement
entropy\cite{Nielsen,Bombelli} or the holographic
principle\cite{'tHooft}.'' They state, ``see, also Ref. 17 for
applying the entanglement entropy in LQG context.'' Thus, we get the
following formula, which is the generalization of Eq.
(\ref{section6ashtekar}):
\begin{widetext}
\begin{equation}
1=\sum_{i} \{(2j_{i}^{u}+1)+(2j_{i}^{d}+1)\}\exp(-8 \pi
\sqrt{\frac{1}{2}\sqrt{2j^{u}_{i}(j^{u}_{i}+1)+2j^{d}_{i}(j^{d}_{i}+1)-j^{t}_{i}(j^{t}_{i}+1)}}/4).
\label{section6pseudo}
\end{equation}
\end{widetext}

In Ref. 5, the authors change the variables
$j^{u}_{i},j^{d}_{i},j^{t}_{i}$ to a new set of integers to
calculate the numerical sum easily. We will not show those details
here. One may easily refer to this construction in their paper.
Another comment on the above formula is that the authors of Ref. 5.
incorrectly used $2j^{u}_{i}+1$ instead of
$(2j^{u}_{i}+1)+(2j^{d}_{i}+1)$ when $j^{u}_{i}$ was equal to
$j^{d}_{i}$. By numerical calculation, we obtained $0.997\cdots$ for
the right-hand side of Eq. (\ref{section6pseudo}). The fact that
this value is so close to 1 suggests that the general area spectrum
obtained by applying our ``newer'' variables in loop quantum gravity
is on the right track. We hope that this very small difference
between $0.997\cdots$ and 1 will be understood in terms of the
effects of the extra dimensions.

\section {Suggestion for a formula for the degeneracy of a single-partition black hole}

The intensity of light with given frequency $\nu$ and with given
radiating object area $A$ in the black body radiation is given by
the following formula, the Planck's law:
\begin{equation}
dI=\frac{2 \pi h \nu^3}{c^2}\frac{A d\nu}{e^{h \nu/k T}-1}.
\label{planck}
\end{equation}
Equivalently, the number of photons produced is given by
\begin{equation}
dn_{photon}=\frac{2 \pi \nu^2}{c^2}\frac{A d\nu}{e^{h \nu/k T}-1}.
\label{number}
\end{equation}

In the case of a black hole, the black hole radiates certain
frequencies of light corresponding to the spectrum of areas. As the
case in Ref. 8, seeing that the following relation between the
energy of the black hole, $h\nu$, and the area spectrum
corresponding to it, $A_{spec}$, must hold is easy:
\begin{equation}
\frac{A_{spec}}{4}=\frac{h\nu}{kT},\label{a4hnukT}
\end{equation}
where $T$ is the temperature of the black hole, and $k$ the
Boltzmann constant. This is so because of the following reason,
which is explained in detail in Ref. 18. If we assume that the
emission of a photon from a black hole is local, it should be
emitted from a single area quantum of the black hole, rather than
from scattered or extended regions of the black hole. One cannot
imagine a photon emitted from two or more places simultaneously.
Therefore, the black hole's area, initially given by $A$, must
decrease by $A_{spec}$ after emitting a photon. Then, at this point,
the black hole's area becomes $A-A_{spec}$, which is smaller than
before; correspondingly, the black hole's energy is also smaller
than before. Therefore, one can see that the energy of the photon
emitted at this moment is exactly given by the energy difference (or
equivalently, the mass difference) between the black hole before and
after the emission of a single photon. With this condition along
with the result of Ref. 7, which is $1/kT=8\pi M$ and $A=16\pi M^2$,
where $M$ is the mass of the black hole, and $A$ the area of the
black hole, one can easily show that this reproduces Eq.
(\ref{a4hnukT}). An easier way of seeing this is the following:

\begin{equation}
T \Delta S= \Delta Q.
\end{equation}

Using $\Delta S=-k \Delta A/4$ and $\Delta Q=-h \nu$, we get Eq.
(\ref{a4hnukT}). Moreover, by plugging Eq. (\ref{a4hnukT}) and the
above results in Ref. 7 into Eq. (\ref{number}), we obtain the
following:
\begin{equation}
dn_{photon}=\frac{A_{spec}^2 dA_{spec}}{2048 \pi^{7/2} \sqrt{A_{BH}}
(e^{A_{spec}/4}-1)}. \label{crucial}
\end{equation}

Considering the fact that the numerator in the above equation is
merely a phase space factor and the general argument presented in
Ref. 8, we first guessed that the denominator might give the number
of states for black- hole areas between $A$ and $A+dA$ as
\begin{equation}
dK(A)= \frac{1}{C} \sqrt{A} (e^{A/4}-1) dA \label{integration}
\end{equation}
for some constant $C$ to be obtained. Here, because we are using
$A=A_{BH}=A_{spec}$ in the above equation, we are considering the
case that the degeneracy of $A_{BH}$ is simply given by the
degeneracy of $A_{spec}$. This is possible only when $A<2 A_{min}$,
where $A_{min}$ is the smallest area spectrum. As in case of $A=2
A_{min}$, $A$ can be partitioned into two unit area sectors, both of
which have the area $A_{min}$. In other words, the above formula is
valid only for such small $A$. Actually, we later found a
justification for Eq. (\ref{integration}) and the fact that it is
valid only for such small $A$. We will present this justification in
Section XI.

In any case, we will obtain the value of $C$ in the next section by
numerical fitting. In other words, we will show that Eq.
(\ref{integration}) works very well. In Sections X-XII, we will
calculate this constant by using another method and show that it
agrees with this value, further supporting our new area spectrum.

\section {Another verification of our area spectrum: the number of states}

The area spectrum we obtained in Section VI is not exact, as there
is a difference between 0.997 and 1. Thus, it may be wrong to use
our non-exact area spectrum to verify Eq. (\ref{integration}).
However, our area spectrum is ``almost'' correct, because the
difference between 0.997 and 1 is very small. Hence, we may as well
use our non-exact area spectrum to verify Eq. (\ref{integration}),
because the exact area spectrum cannot be obtained at this point.

To this end, we integrate both hand-sides of (\ref{integration});
then, we get:
\begin{equation}
C(A)=\frac{I(A)}{K(A)}, \label{C}
\end{equation}
where $K(A)$ is the number of states for ab area equal to, or below,
$A$, and $I(A)$ is given by
\begin{equation}
I(A)=\int_{A_{cut}}^A \sqrt{A'} (e^{A'/4}-1) dA'.\label{I}
\end{equation}
In an ideal case, when our area spectrum totally respects Eq.
(\ref{integration}), $C(A)$ is a constant that does not depend on
$A$. Here, $A_{cut}$ is a somewhat arbitrary cut-off for the
integration as there is a non-zero minimum allowed value for
$A_{spec}$ in Eq. (\ref{crucial}) and, therefore, $A$ in Eq.
(\ref{integration}). One may naively guess that $A_{cut}$ should be
this non-zero minimum allowed value for $A_{spec}$, which we called
$A_{min}$ in the last section, but that would lead to a nonsensical
result as $C(A_{min})$ would be zero, because $I(A_{min})$ would be
zero, in such a case. Therefore, $A_{cut}$ should be slightly
smaller than $A_{min}$, but not too much. Therefore, say, we suggest
choosing conveniently the difference between $A_{cut}$, the starting
point of the integration, and $A_{min}$, the smallest area spectrum,
to be the difference between the smallest area spectrum and the
second smallest area spectrum. Then, we can set $A_{cut}$ to be
given by
\begin{equation}
17.8-A_{cut}=21.1-17.8\label{Acut}
\end{equation}
as 17.8 is the smallest area value and 21.1 is the second smallest
area value. You can find these values in Table 1.

Now, let me explain Table 1. For convenience, we defined the
variable $y$ as
\begin{equation}
y=2j^{u}_{i}(j^{u}_{i}+1)+2j^{d}_{i}(j^{d}_{i}+1)-j^{t}_{i}(j^{t}_{i}+1).\label{c}
\end{equation}
Then, we can easily see the relation
\begin{equation}
A(y)=8 \pi \sqrt{\frac{1}{2} \sqrt{y}}.\label{A}
\end{equation}
In other words, $y$ is a positive integer, which labels the area
spectrum. Putting everything together and using Mathematica, we
obtain Table 1. Here, we have calculated up to $y=15$ because for
$y$ bigger than, or equal to, 16, the black-hole area degeneracy can
no longer be equal to the unit area spectrum degeneracy. For
example, when $y=16$, we have $A(16)=8\pi \sqrt 2$, which can also
be partitioned into two unit area sectors, both of which have the
area $A(1)=4\pi\sqrt 2$. Recall that we have actually mentioned this
at the end of Section VIII. \textit{i.e.}, $A<2 A_{min}$. We also
want to remark that we have done the calculation for
multi-partitioned black holes \textit{i.e.}, $A\geq 2 A_{min}$, or
equivalently $y\geq 16$ and confirmed that the fitting with Eq.
(\ref{integration}) did not work at all, as expected, beginning from
$A=2A_{min}$, \textit{i.e.} $y=16$.

\begin{table}[t!]
\caption[0]{Values of C(A).}
\begin{ruledtabular}
\begin{tabular}{ccccc}
y & A & K(A)& I(A)& C(A) \\ \hline
1 & 17.8 & 4 & 767.4& 191.8 \\
2 & 21.1 & 14 & 2740& 195.7 \\
3 & 23.4 & 32 & 5552& 173.5 \\
4 & 25.1 & 50 & 9276& 185.5 \\
5 & 26.6 & 72 & 14000& 194.4 \\
6 & 27.8 & 110 & 19814& 180.1 \\
7 & 28.9 & 154 & 26817& 174.1 \\
8 & 29.9 & 204 & 35109& 172.1 \\
9 & 30.8 & 262 & 44797& 171.0 \\
10 & 31.6 & 326 & 55990& 171.7 \\
11 & 32.4 & 388 & 68803& 177.3 \\
12 & 33.1 & 474 & 83353& 175.8 \\
13 & 33.7 & 584 & 99761& 170.8 \\
14 & 34.4 & 684 & 118155& 172.7 \\
15 & 35.0 & 804 & 138664& 172.5
\end{tabular}
\end{ruledtabular}
\begin{flushleft} \end{flushleft}
\end{table}


From the table, one can find that $C(A)$ does not strongly depend on
$A$ and that the biggest value for $C(A)$ in our result is 195.7,
deviating from the ``right value'' of $C$, \textit{i.e.}, the value
of $C$ for large $A$ by only about 13 percent. As $K(A)$ at this
biggest value of $C(A)$ is only 14, it necessarily has a big
``statistical'' variation. Therefore, 195.7 is not a big deviation.

\section {Motivated by Bekenstein and Mukhanov}
In this section and the next two sections, we derive $C$ by using
another method and present a reason behind Eq. (\ref{integration}).
To this end, we will present formulae motivated by Ref. 8. Let us
say that $j_{\Delta t}$ denotes the average number of emitted photon
from a black hole during the time $\Delta t$. If we assume that
$\Delta t$ is sufficiently small, $j_{\Delta t}$ is proportional to
$\Delta t$. Therefore, we can write
\begin{equation}
j_{\Delta t}=\frac{\Delta t}{\tau}
\end{equation}
for some $\tau$ to be determined.

On the other hand, the decrease in the mass of the black hole is the
average energy of the emitted photon multiplied by the average
number of emitted photons. Therefore, we have
\begin{equation}
\Delta M=-\frac{\int dn_{photon}h\nu }{\int dn_{photon}}\frac{\Delta
t}{\tau},\label{19}
\end{equation}
where $dn_{photon}$ is the number of photons with a given frequency
between $\nu$ and $\nu+d\nu$ emitted during unit time, which is
given by Eq. (\ref{number}), which we reproduce here for
convenience:
\begin{equation}
dn_{photon}=\frac{2\pi
\nu^2}{c^2}\frac{A_{BH}d\nu}{e^{h\nu/kT}-1}.\label{dnphoton}
\end{equation}

Now, observe the condition
\begin{equation}
\frac{\Delta M}{\Delta t}=-\int dn_{photon}h\nu,
\end{equation}
which is obvious because $h\nu$ is the energy of a single photon
with frequency $\nu$. By plugging this formula into Eq. (\ref{19}),
we obtain:
\begin{equation}
\frac{1}{\tau}=\int dn_{photon}.\label{tau}
\end{equation}
Now, let us say that during the time $\Delta t$, $x_{a, \Delta t}$
number of photons that correspond to a decrease in the black hole
area by ``$a$'' are emitted. Apparently, we have
\begin{equation}
1=\sum_y \frac{x_{a(y), \Delta t}}{j_{\Delta t}} \label{23}
\end{equation}
because during the time $\Delta t$, the black hole's area can
decrease by any area spectrum ``$a$''. Here, we have used the
notation of Eq. (\ref{A}) and called the area spectrum $a$ instead
of $A_{spec}$.

\section{Justification for our formula for the degeneracy of a single-partition black hole}

Given the above, how can we relate Eq. (\ref{23}) with Eq.
(\ref{dnphoton})? From the definition of $x_{a,\Delta t}$, we can
relate them by the following formula
\begin{equation}
x_{a, \Delta t} =\Delta n_{photon}\Delta t. \label{tying}
\end{equation}
Here, $\Delta n_{photon}$ is just a discrete version of Eq.
(\ref{dnphoton}) given by
\begin{equation}
\Delta n_{photon}=\frac{2\pi
\nu^2}{c^2}\frac{A_{BH}\Delta\nu}{e^{h\nu/kT}-1},\label{Deltanphoton}
\end{equation}
where, naturally, $a$ and $\nu$ are related by Eq. (\ref{a4hnukT}).
An easy way to check that Eq. (\ref{tying}) is correct is summing it
over $y$. Then, we get Eq. (\ref{tau}).

Now, let us focus on the right-hand side of Eq. (\ref{tying}). To
this end, let us digress to the topic of the density of states. For
a cube with size $L\times L\times L$, we have the following formulae
for the momentum:
\begin{eqnarray}
&& p_x=\frac{h n_x}{2L},\nonumber\\
&& p_y=\frac{h n_y}{2L},\nonumber\\
&& p_z=\frac{h n_z}{2L}.
\end{eqnarray}
By relating the momentum and the frequency of photons as $p=h\nu/c$
and by obtaining the density of states by using a standard
procedure, we get
\begin{equation}
\nu^2 \Delta\nu=\frac{c^3}{h^3}p^2\Delta p=\frac{c^3}{8L^3}(n^2
\Delta n).\label{fermi}
\end{equation}

A natural choice for $\Delta n$ is
\begin{equation}
2(\frac{\pi}{2}n^2 \Delta n)=1,\label{1}
\end{equation}
where $\frac{\pi}{2} n^2 \Delta n$ comes from the surface area of
one-eighth of a sphere, as $n_x$, $n_y$, $n_z$ are positive and the
factor 2 comes from the two polarizations of photons. In other
words, this is a standard procedure for the density of states.

Substituting Eq. (\ref{1}) into Eq. (\ref{fermi}), then to Eq.
(\ref{Deltanphoton}), and then to Eq. (\ref{tying}), Eq.
(\ref{tying}) becomes
\begin{equation}
\Delta t/\tau \frac{x_{a, \Delta t}}{\Delta
t/\tau}=\frac{c}{4L^3}\frac{A_{BH}}{e^{a/4}-1} \Delta
t.\label{righthandside}
\end{equation}
As $L^3$ is the volume of the cube, we can regard it as the volume
of the black hole, even though the shape of the black hole is not
rectangular, but spherical; we would have obtained the same result
if we had considered the density of states of photons confined in a
sphere, even though the actual calculation would have been different
and more complicated. All that matters is that we recover the volume
factor for $L^3$. Given this, using the relations
\begin{eqnarray}
&& A_{BH}=4\pi r^2,\nonumber\\
&& L^3=\frac{4\pi r^3}{3}
\end{eqnarray}
we get
\begin{equation}
L^3=\frac{A_{BH}\sqrt{A_{BH}}}{6\sqrt{\pi}}.\label{Lcube}
\end{equation}
Now, we can plug Eq. (\ref{Lcube}) into Eq. (\ref{righthandside}),
and get
\begin{equation}
x_{a,\Delta t}=\frac{3\sqrt{\pi}
c}{2\sqrt{A_{BH}}}\frac{1}{e^{a/4}-1}\Delta t.
\end{equation}

Given the above, going back to Bekenstein and Mukhanov's work is
worthwhile\cite{Bekenstein}. They wrote, ``the probability for the
jump (\textit{i.e.} emission) must be proportional to the final
level's degeneracy.'' In other words, they are saying that
$x_{a,\Delta t}/j_{\Delta t}$ must be proportional to the degeneracy
of the black hole after emission divided by the degeneracy of the
black hole before emission. Denoting $W(A_{BH})$ for the degeneracy
of a black hole with area $A_{BH}$, we have
\begin{equation}
\frac{W(A_{BH}-a)}{W(A_{BH})}\propto \frac{\tau}{\sqrt
{A_{BH}}}\frac{1}{e^{a/4}-1}.\label{W}
\end{equation}
Now, if a black hole is made of a single partition with area $A$, it
can only emit a photon associated with the area spectrum $a=A$ and
not others. In other words, we have
\begin{equation}
A_{BH}=a=A,\qquad\mathrm{where~} A<2A_{min}\label{whereA<2min}
\end{equation}
Plugging Eq. (\ref{whereA<2min}) into Eq. (\ref{W}), we get
\begin{equation}
\frac{W(0)}{W(A)}\propto \frac{\tau}{\sqrt {A}}\frac{1}{e^{A/4}-1},
\end{equation}
which is exactly Eq. (\ref{integration}) if $\tau$ is universal for
single-partition black holes.

\section {Tying the number of degeneracy into the Hawking radiation spectrum}
We can also evaluate the left-hand side of Eq.
(\ref{righthandside}). However, we must be careful when we calculate
$\tau$ there. As was shown earlier, no light is emitted below a
certain frequency in the Hawking radiation. Considering this, $\tau$
in the Eq. (\ref{tau}) is given by the following expression:

\begin{equation}
\frac{1}{\tau}=A_{BH}\frac{2\pi}{c^2}\int_{\pi\sqrt{2}}^{\infty}\frac{u^2
du}{e^u-1}(\frac{kT}{h})^3.\label{1tau}
\end{equation}
Namely, the strange factor $\pi\sqrt{2}$ denotes the fact that
$h\nu_{min}$, the minimum energy of a photon emitted from the black
hole with temperature $T$, is given by
\begin{equation}
h\nu_{min}=\pi\sqrt{2} kT=\frac{A_{min}}{4} kT
\end{equation}
In other words, we approximate our new Hawking radiation spectrum as
being continuous, but with a minimum emitted photon energy. This is
reasonable from the following reason. With the notation of Eq.
(\ref{A}), the difference between $A(0)$ and $A(1)$ is big while the
difference between $A(n)$ and $A(n+1)$ for a non-negative integer
$n$ is quite small, implying that the spectrum of the emitted
frequency is ``dense'' enough to be considered ``continuous'' in
this range.

Using the fact that $kT=1/(8\pi M)$ and $A_{BH}=16\pi M^2$ and tying
everything together, we obtain
\begin{equation}
\frac{x_{a(y), \Delta t}}{j_{\Delta
t}}=\frac{6\pi}{\alpha(e^{a/4}-1)},\label{ff}
\end{equation}
where $\alpha$ is given by
\begin{equation}
\int_{\pi\sqrt{2}}^{\infty}\frac{u^2 du}{e^u-1}=0.36193...
\end{equation}
Recalling Eq. (\ref{23}), we can take the summation with respect to
$a$ on both sides of Eq. (\ref{ff}). This yields
\begin{equation}
1=\sum_y \frac{6\pi}{\alpha(e^{a(y)/4}-1)}.\label{sumy}
\end{equation}
From the above equation, let us derive another equivalent equation
that should be satisfied, but contains $C$ so that we can obtain its
value in a new way and compare it to the one obtained in Section IX.

First, notice the following:
\begin{eqnarray}
\sum_a&=&\sum_y \left(K(a(y))-K(a(y-1))\right)\nn \\ &=&\sum_y
\left(K(a(y+\frac 12))-K(a(y-\frac 12))\right),\label{sumysuma}
\end{eqnarray}
where $\sum_a$ is the summation that takes into account the
degeneracy. This formula is natural, because there are
$K(a(y))-K(a(y-1))$ number of degeneracies for an area spectrum
``$a$'' and because $K(a(y))=K(a(y+\frac 12))$ for integer $y$;
$K(a(y))$ increases only discretely. To make the notation $\sum_a$
easier to understand, let us give an example:
\begin{eqnarray}
 1 &=&\sum_a e^{-a/4}\nn\\
 &=&\sum_y \left(K(a(y+\frac 12))-K(a(y-\frac 12))\right)
e^{-a(y)/4}.\nn\\
&&
\end{eqnarray}

Now, we need to express $K(a(y+\frac 12))-K(a(y-\frac 12))$ by using
$C$. To this end, from Eq. (\ref{A}), we have
\begin{equation}
a=4\sqrt{2}\pi y^{1/4}.
\end{equation}
Also, let us use the notation
\begin{eqnarray}
\Delta a&=&(a+\frac{\Delta a}{2})-(a-\frac{\Delta a}{2})= a(y+\frac
12)-a(y-\frac 12)\nn \\&=&4\sqrt{2}\pi (y+\frac
12)^{1/4}-4\sqrt{2}\pi (y-\frac 12)^{1/4}.
\end{eqnarray}
Therefore, for large $a$, we get
\begin{equation}
\Delta a=\frac{256\pi^4}{a^3}.\label{Deltaa}
\end{equation}

Given this, from Eqs.~(\ref{C}) and (\ref{I}), we have
\begin{eqnarray}
&& \hs K(a+\frac{\Delta a}{2})-K(a-\frac{\Delta a}{2}) \nn
\\ && =\frac{I(a+\Delta a/2)-I(a-\Delta
a/2)}{C}\nn\\ && =\frac{\sqrt{a}(e^{a/4}-1)\Delta a}{C}\label{N}
\end{eqnarray}
Now, we use Eq. (\ref{sumysuma}) to re-express Eq. (\ref{sumy}) and
obtain
\begin{equation}
1=\sum_a \frac{6\pi}{\alpha(e^{a/4}-1)}\frac{1}{K(a+\Delta
a/2)-K(a-\Delta a/2)}.~~~
\end{equation}
Plugging Eq.~(\ref{Deltaa}) and Eq.~(\ref{N}) into the above
equation, we obtain
\begin{equation}
1=\frac{3C}{128\pi^3\alpha}\sum_a \frac{a^{5/2}}{(e^{a/4}-1)^2}.
\end{equation}
Using
\begin{equation}
\sum_a \frac{a^{5/2}}{(e^{a/4}-1)^2}=2.7697...,
\end{equation}
we obtain
\begin{equation}
C=172.87...,
\end{equation}
which agrees with the value 172 $\sim$ 173 that we obtained in
Section IX.

\section {The naive black hole degeneracy}
In loop quantum gravity, the entropy of a black hole is calculated
in two steps. First, as Rovelli proposed, one counts the number of
ways in which the area of the black hole can be expressed as the sum
of unit areas. Taking the log of that number (and upon fixing
Immirzi parameter), we get $A/4$ (in the leading order). Second, one
takes into account the projection constraint or the $SU(2)$
invariant subspace constraint. This gives the logarithmic
corrections.

In Ref. 19, one of us named the black hole degeneracy calculated
before taking into account the projection constraint or the $SU(2)$
invariant subspace constraint as the ``naive'' black hole
degeneracy. There, the naive black hole degeneracy was shown to be
close to $e^{A/4}$ and does not have any logarithmic corrections;
nevertheless, it does have some sub-leading corrections. These
corrections are calculated in the paper.

Here, we want to argue that the black hole degeneracy is given by
the naive black hole degeneracy if the area spectrum on the black
hole's horizon is given by the general area spectrum (either in its
newer variables version or in its original Ashtekar variables
version a la Tanaka-Tamaki) rather than by the isolated horizon
framework. This is so because the general area spectrum is already a
$SU(2)$ invariant, because one takes into account the intertwiner in
the spin network state when one calculates the area spectrum.
Remember that the intertwiner makes the spin network $SU(2)$ gauge
invariant.

\section {Discussion and Conclusions}
In the first part of our paper we introduced ``newer'' variables,
and by quantizing them, we applied them to calculating the spectrum
of area. We obtained that the spectrum of area is the square root of
the spectrum of area predicted by Ashtekar's variable theory. By
using this result for the area spectrum, we showed that our result
``almost correctly'' predicted the Bekenstein-Hawking entropy
formula. This is very remarkable because no one, in the current
framework of loop quantum gravity, has succeeded yet in predicting
the Bekenstein-Hawking entropy formula without adjusting the Immirzi
parameter. Of course, this is because the Immirzi parameter, which
is explained in Ref. 20, plays an indispensable role in the current
framework of loop quantum gravity. Some have also noted that the
same choice of the Immirzi parameter yields the Bekenstein-Hawking
entropy universally for various kinds of black holes, as if this
could be evidence for the concept of the Immirzi
parameter\cite{ABCK}. However, this does not say much as the leading
term to the black hole entropy, given by the Bekenstein-Hawking
entropy formula, depends solely on the area of the black hole and
the area spectrum, which is universal for all kinds of black holes.
Therefore, a consistent check seems to be lacking, as no one has
computed the value of the Immirzi parameter by ways other than
adjusting it to make the Bekenstein-Hawking entropy formula hold.
Moreover, the Immirzi parameter is ``ugly,'' In its presence, we
have an ugly and complicated term in the action whereas without it
(or if the Immirzi parameter is $i$), we have a simple and beautiful
Einstein-Hilbert action. String theory can derive the
Einstein-Hilbert action, but imagining how the ugly term due to the
Immirzi parameter would be derived is difficult.

We also conjectured that the very small discrepancy between $0.997
\cdots$ and 1 was due to the extra dimensions that seem to modify
the area spectrum. Thus, we hope that our theory will be a very
useful tool for probing extra dimensions, as it seems to give a very
strong constraint on its size. We hope that future endeavors
determine it or at least its rough length scale. For example, one
may check whether Arkani-hamed, Dimopoulos and Dvali's
proposal\cite{Nima} that new dimensions are at millimeters is
consistent with our conjecture, or one may also check the
Randall-Sundrum model\cite{Randall}. Nevertheless, the most ideal
case would be if Klein's prediction\cite{Klein} that the size of the
5th dimension (in case it's a circle) is $8.428\times 10^{-33}m$ is
reconfirmed by resolving our conjecture. In other words, he argued
that the size of the 5th dimension is given by

\begin{equation}
\frac{(4\pi)^{3/2}}{\sqrt{\alpha}}l_p,
\end{equation}
where $\alpha$ is the fine structure constant, and $l_p$ is Planck's
length. (This is the case when the electric charges are integer
multiples of ``$e$'' the electronic charge. If we consider quarks
whose electric charges are given by $e/3$, the size of
extra-dimension should be tripled.)

At any rate, if our conjecture is resolved, it would be a good
solution to the fine-tuning problem and the hierarchy problem. This
is so because 0.997 is very close to 1, even though no fine-tuning
was done. Moreover, if the extra dimension scale is obtained by the
difference 0.003, the hierarchy problem will be solved, as we will
have a natural explanation for it. Even though it is too early to
judge the situation, one may guess that \textit{all} kinds of
fine-tuning problems and hierarchy problems will be solved in this
manner.

In the second phase of our paper, we numerically showed that our
area spectrum correctly reproduces the degeneracy of a
single-partition black hole derived by using the method inspired by
Bekenstein and Mukhanov's work. Remarkably, using two different
methods, we obtained the same value of $C$ that related the area
spectrum to the degeneracy of a single-partition black hole,
strongly supporting our area spectrum and its consistency.

In an appendix, we explicitly checked that this consistency that the
two different methods yielded the same value of $C$ did not hold in
the cases of area spectra based on the Ashtekar variables with the
corresponding $C$ defined appropriately. Actually, the situation is
much worse than that. $C$ is not even a constant if we determine it
by using the first method, as the fittings do not work at all. In
other words, the area spectra based on the Ashtekar variables do not
reproduce the degeneracy of a single-partition black hole.

Incidentally, our paper has weaknesses. We have an additional $i$
factor in the Poisson bracket between certain combinations of
``newer'' variables. This is troublesome because it implies that the
area spectrum is not real. Also, we need to understand why $\tau$
seems to be universal for single-partition black holes. Certainly,
in such a case, to calculate $\tau$, we cannot use Eq.
(\ref{dnphoton}), which is not valid for such a black hole. How a
single-partition black hole Hawking radiates is a problem that needs
to be solved and understood.

Finally, our new area spectrum gives testable predictions should the
Hawking radiation be observed at the LHC. As a gap exists between
zero and the smallest value of the area allowed, no photons are
emitted below a certain frequency that we calculate to be $h \nu/ k
T\approx4.44$. Moreover, the black hole evaporates at a certain
slower rate than expected by a naive application of Hawking's theory
precisely because of this. Furthermore, if we are lucky and the
measurement at the LHC is sensitive enough, we may confirm
sub-leading corrections to the Bekenstein-Hawking entropy. In Ref.
18, suggestions on how to do this are given. These corrections seem
to be especially important because the black holes that many hope
will be created at the LHC will not be too big to ignore the
sub-leading corrections; indeed, many call them mini-black holes.

We hope that our predictions will be confirmed at the LHC. If we are
luckier, Klein's prediction may be reconfirmed by solving our
conjecture before the LHC detects any black-hole productions.

\begin{acknowledgments}

BK calculated $K(A)$ in Tables 1, 3 and 4 by using java. YY did the
rest of the work. The work of YY was supported by National Research
Foundation of Korea (NRF) grants 2012R1A1B3001085 and
2012R1A2A2A02046739. YY thanks Jong-Hyun Baek for bringing his
attention to the fact that hodge dual always includes an extra
$\sqrt{g}$ factor (\textit{i.e.}, Levi-Civita tensor) for the
spacetime variables.

\end{acknowledgments}

\appendix
\section{Tentative falsification of the area spectrum
based on Ashtekar variables}


In this appendix, we show that our second piece of evidence for our
new area spectrum based on ``newer'' variables, presented in Section
VIII and IX, does not work in the cases of the area spectra based on
the traditional Ashtekar variables. We call it a ``tentative''
falsification as our derivation of Eq. (\ref{integration}) in
Section XI needs a further justification that $\tau$ is universal
for single-partition black holes. Once a good reasoning for this is
found, the result of this appendix must be regarded as a genuine
falsification.

Here, we will consider the three area spectra based on Ashtekar
variables: the isolated horizon framework, the original
Tanaka-Tamaki scenario and the modified Tanaka-Tamaki scenario. In
the isolated horizon framework, the relevant area spectrum is given
by
\begin{equation}
A=8\pi\gamma \sum_{i} \sqrt{j_i(j_i+1)},
\end{equation}
where $\gamma$ is the Immirzi parameter and $j_i$ are non-negative
half-integers. For a given $j_i$, the above area spectrum has a
degeneracy of $(2j_i+1)$. In this case, the Immirzi parameter is
equal to $0.274067\cdots$\cite{Ghosh}. Now, we present Table 2,
which is the isolated horizon framework version of Table 1. Here, we
used the corresponding $A_{cut}$ as in Eq. (\ref{Acut}). We clearly
see that C(A) does not converge. We want to mention that we have
unnecessarily included the cases of big enough $A$s whose
degeneracies are not simply equal to the degeneracy of unit area
spectrum to show that $C$ is not a constant for high $A$. Also, we
have unnecessarily calculated the relevant $C$ by using the methods
of section X and XI. We obtained 9.62$\cdots$ which is far from the
$C(A)$ presented in Table 2.

\begin{table}[t!]
\caption[0]{Isolated horizon.}
\begin{ruledtabular}
\begin{tabular}{ccccc}
j & A & K(A)& I(A)& C(A) \\ \hline
0.5 & 6.0 & 2 & 15.0 & 7.5 \\
1 & 9.7 & 5 & 83.9& 16.8 \\
1.5 & 13.3 & 9 & 300& 33.4 \\
2 & 16.9 & 14 & 911& 62.1 \\
2.5 & 20.4 & 20 & 2548& 127.4 \\
3 & 23.9 & 27 & 6818& 252.5 \\
3.5 & 27.3 & 35 & 17765& 507.6 \\
4 & 30.8 & 44 & 45504& 1034 \\
4.5 & 34.3 & 54 & 115195& 2133 \\
5 & 37.7 & 65 & 289140& 4448
\end{tabular}
\end{ruledtabular}
\begin{flushleft} \end{flushleft}
\end{table}

%

\begin{table}[t!]
\caption[0]{Original Tanaka-Tamaki scenario.}
\begin{ruledtabular}
\begin{tabular}{ccccc}
y & A & K(A)& I(A)& C(A) \\ \hline
1 & 7.3 & 2 & 24.7& 12.4 \\
2 & 10.3 & 9 & 99.4& 11.0 \\
3 & 12.6 & 21 & 232& 11.1 \\
4 & 14.6 & 34 & 444& 13.1 \\
5 & 16.3 & 50 & 761& 15.2 \\
6 & 17.9 & 78 & 1219& 15.6 \\
7 & 19.3 & 114 & 1858& 16.3 \\
8 & 20.6 & 152 & 2730& 18.0 \\
9 & 21.9 & 196 & 3899& 19.9 \\
10 & 23.1 & 249 & 5441& 21.9
\end{tabular}
\end{ruledtabular}
\begin{flushleft} \end{flushleft}
\end{table}


\begin{table}[t!]
\caption[0]{Modified Tanaka-Tamaki scenario.}
\begin{ruledtabular}
\begin{tabular}{ccccc}
y & A & K(A)& I(A)& C(A) \\ \hline
1 & 10.7 & 4 & 104 & 26.0 \\
2 & 15.2 & 14 & 524& 37.4 \\
3 & 18.6 & 32 & 1501& 46.9 \\
4 & 21.5 & 50 & 3462& 69.2 \\
5 & 24.0 & 72 & 7072& 98.2 \\
6 & 26.3 & 110 & 13332& 121.2 \\
7 & 28.4 & 154 & 23714& 154.0 \\
8 & 30.4 & 204 & 40345& 197.8 \\
9 & 32.2 & 262 & 66241& 252.8 \\
10 & 33.9 & 326 & 105618& 324.0
\end{tabular}
\end{ruledtabular}
\begin{flushleft} \end{flushleft}
\end{table}


Next, we consider the original Tanaka-Tamaki scenario. As mentioned,
Tanaka and Tamaki\cite{Tanaka} used the wrong degeneracy.
Nevertheless, we unnecessarily show that our numerical evidence does
not work for their area spectrum. As was the case with our analysis
of the isolated horizon framework, the relevant $A_{cut}$ is used,
and the calculation in the case of high $A$s is included to show
that $C$ is not constant for high $A$. The results are shown in
Table 2, where $A=8\pi\gamma\sqrt y$. We clearly see that $C$ is not
constant. For your information, we also obtained $C=41.37\cdots$ by
using the methods of section X-XII.

Lastly, we consider the modified Tanaka-Tamaki scenario. In this
scenario, we corrected the wrong degeneracy obtained by Tanaka and
Tamaki. Otherwise, the case is similar to the original one. In
particular, every comment on the calculation setting of the original
Tanaka-Tamaki scenario applies to the modified Tanaka-Tamaki
scenario. Again, we see that $C$ is not constant. For information,
we obtained $C$ to be $154.16\cdots$ by using the methods of section
X-XII.

\section{Kaluza-Klein Theory and Its Extension}

In this appendix, we discuss a possible direction to confirm our
conjecture that the consideration of an extra dimension modifies the
area spectrum in such a way that the Bekenstein-Hawking entropy is
satisfied. A very well-known scenario for an extra dimension is the
Kaluza-Klein theory\cite{Klein}. Klein suggested a metric for his
five-dimensional theory and showed that the metric led to the
unification of gravity and Maxwell's electromagnetic field. Let us
review how that works. We closely follow Ref. 25.

Imagine that we are in five dimensions with metric components
$G^{(5)}_{MN}$, $M, N=0,1,2,3,4$ and that the spacetime actually has
topology $\textbf{R}^4 \times S^1$, and so has one compact
direction. There, we will have the usual four-dimensional
coordinates on $\textbf{R}^4$, $x^{\mu}$ ($\mu, \nu=0,1,2,3$), and a
periodic coordinate

\begin{equation}
x^4=x^4+2 \pi R,
\end{equation}
where $2 \pi R$ is the size of the extra dimension.

Now, under the five-dimensional coordinate transformation
$x'^M=x^M+\epsilon^M(x)$, the five-dimensional metric transforms as
follows:
\begin{equation}
G^{(5)'}_{MN}=G^{(5)}_{MN}-\partial_M
\epsilon_N-\partial_N\epsilon_M. \label{GMN}
\end{equation}
Given this, let us assume that the metric does not depend on the
periodic coordinate $x^4$. Then, we immediately see the following:
\begin{equation}
\epsilon^\nu=\epsilon^\nu(x^\mu),
\end{equation}
\begin{equation}
\epsilon^4=\epsilon^4(x^\mu),
\end{equation}
which means
\begin{equation}
x^{\mu'}=\psi^\mu(x^0,x^1,x^2,x^3),
\end{equation}
\begin{equation}
x^{4'}=x^4+\epsilon^4(x^0,x^1,x^2,x^3).
\end{equation}
They have obvious physical interpretations. The first one is the
usual four-dimensional diffeomorphism invariance. The second one is
an $x^\mu$-dependant isometry(rotation) of the circle.

Then, from Eq. (\ref{GMN}), $G^{(5)}_{44}$ is invariant, and we also
have

\begin{equation}
G^{(5)'}_{\mu 4}=G^{(5)}_{\mu 4}-\partial_{\mu} \epsilon_4.
\end{equation}
However, from the four-dimensional point of view, $G^{(5)}_{\mu 4}$
is a vector that is proportional to what we will call $A_\mu$, so
the above equation is simply a $U(1)$ gauge transformation for the
electromagnetic potential $A_{\mu}'=A_{\mu}-\partial_\mu \Lambda$.
Thus, the $U(1)$ of electromagnetism can be thought of as resulting
from compactifying gravity, the gauge field being an internal
component of the metric. If $G_{44}=1$ is assumed, these
considerations lead to the following metric:

\begin{equation}
ds^2=G^{(5)}_{MN}dx^M dx^N=G^{(4)}_{\mu\nu}
dx^{\mu}dx^{\nu}+(dx^4+A_\mu dx^{\mu})^2,\label{G4}
\end{equation}
Given this, the five-dimensional Ricci scalar can be re-expressed in
terms of the four-dimensional one and the electromagnetic field
tensor as follows:
\begin{equation}
R^{(5)}=R^{(4)}-\frac{1}{4} F_{\mu\nu} F^{\mu\nu},
\end{equation}
where $F_{\mu\nu}=\partial_\mu A_{\nu}-\partial_\nu A_{\mu}$ as
usual.

Now, we have
\begin{eqnarray}
S&=&\frac{1}{16\pi G^N_{(5)}}\int d^5x (-G_{(5)})^{1/2} R^{(5)}
\\
&=&\frac{1}{16\pi G^N_{(4)}}\int d^4x (-G_{(4)})^{1/2}
(R^{(4)}-\frac{1}{4} F_{\mu\nu} F^{\mu\nu})\nn\\&&
\end{eqnarray}
Therefore, the Einstein-Hilbert action in five-dimensional
Kaluza-Klein theory reproduces the Einstein-Hilbert action in
four-dimensional theory and the Maxwell Lagrangian, which means a
unification of gravity and electromagnetism. Also, from the above
formulae, clearly, we have the following relation between the
5-dimensional Newton's constant and the 4-dimensional Newton's
constant:
\begin{equation}
\frac{2\pi R}{G^N_{(5)}}=\frac{1}{G^N_{(4)}},
\end{equation}
where $2 \pi R$ is the size of the extra-dimension, as stated
before.

In Ref. 26, Einstein and Bergmann went beyond the Kaluza-Klein
theory. Dropping the condition that the metric be independent of the
periodic coordinate, they went on to introduce a ``new'' covariant
derivative that involved differentiation with respect to $x^0$ other
than the usual Levi-Civita connection. (From now on, the compact
direction is not $x^4$, but $x^0$.) Stepping further, they defined a
``new'' Riemann tensor. Then, they wrote all the possible terms for
the action, which included two more terms other than the
Einstein-Hilbert action (which is given in terms of a ``new''
Riemann tensor) and the Maxwell Lagrangian (see Eqs. (32) and (33)
on page 694). Here, $A_{mn}$ means the $m$ and $n$th component of
the electromagnetic tensor, which is usually called $F_{mn}$.
Nevertheless, they failed to get the relative coefficient between
these terms. In other words, their theory has free parameters. For
convenience, their formulae are reproduced here and from now on $g$
means $G^{(4)}$ in the language of Eq. (\ref{G4}).
\begin{equation}
H_1=R^i_{klm}\delta^l_i g^{km}=R_{km}g^{km}=R,
\end{equation}
\begin{equation}
H_2=A_{mn}A^{mn},
\end{equation}
\begin{equation}
H_3=g^{mn} {}_{,0}g_{mn,0},
\end{equation}
\begin{equation}
H_4=g^{mn}g_{mn,0}g^{rs}g_{rs,0},
\end{equation}

\begin{eqnarray}
S&=&\int dx^0 dx^1 dx^2 dx^3 dx^4 \sqrt{-g}\nn \\
&&\times(\alpha_1 H_1+\alpha_2 H_2 +\alpha_3 H_3 +\alpha_4 H_4).
\end{eqnarray}

However, in addition to the usual differential equations, the
variation of the above action leads to integro-differential
equations because of the periodicity of the compact coordinate
$x^0$. In other words,
\begin{eqnarray}
0&=&\delta S=\int_{0}^{2 \pi R} dx^0 \delta(\int dx^1 dx^2 dx^3
dx^4\sqrt{-g}\nn \\ &&\times(\alpha_1 H_1+\alpha_2 H_2+\alpha_3
H_3+\alpha_4 H_4)).\label{EB}
\end{eqnarray}
Luckily, in Ref. 27, Einstein \textit{et al.} succeeded in obtaining
pure differential equations which are free of integration. Moreover,
using clever ideas and ``customized'' Bianchi identities that one
gets when the metric depends on $x^0$, they went on to derive exact
field equations without the arbitrariness of the above free
parameters. Please check Eqs. (37) and (38) in their paper, which
are reproduced here for convenience:
\begin{widetext}
\begin{eqnarray}
G^{ik}&\equiv&\frac{1}{2}(R^{ik}+R^{ki}-g^{ik}R)+\frac{1}{2}(A^{is}A^k_s-\frac{1}{4}g^{ik}A^{rs}A_{rs})-g^{ik}[\frac{3}{8}g^{rs} {}_{,0}g_{rs,0}+\frac{1}{8}(g^{rs}g_{rs,0})^2]\nonumber\\
&&+\frac{1}{2}g^{im} {}_{,0}g^{ks}g_{sm,0}-\frac{1}{4}g^{ik}
{}_{,0}(g^{rs}g_{rs,0})+\frac{1}{2}(g^{ir}g^{ks}-g^{ik}g^{rs})g_{rs,00}=0,
\end{eqnarray}
\end{widetext}
\begin{equation}
G^i\equiv A^{im}
{}_{;m}-g^{im}g^{rs}g_{mr,0;s}+g^{im}g^{rs}g_{rs,0;m}=0.
\end{equation}
They compared their results with the earlier result obtained by
Einstein and Bergmann\cite{EB} and noted that their theory gave
\begin{equation}
\alpha_1=1
~~~~\alpha_2=\frac{1}{4}~~~~\alpha_3=\alpha_4=-\frac{1}{4}
\end{equation}
and that in their language, the differential equations derived from
Eq. (\ref{EB}) correspond to
\begin{equation}
G^{ik}=0
\end{equation}
while the integro-differential equations derived from Eq. (\ref{EB})
correspond to
\begin{equation}
\int_0^{2 \pi R} dx^0 \sqrt{-g} G^i=0,
\end{equation}
Thus, the field equations of Einstein \textit{et al.} are
advantageous because their field equations are stronger than the
earlier ones of Einstein and Bergmann and are free of free
parameters.

\end{document}